# Reduction of gas bubbles and improved critical current density in Bi-2212 round wire by swaging

Jianyi Jiang, Hanping Miao, Yibing Huang, Seung Hong, Jeff A. Parrell, Christian Scheuerlein, Marco Di Michiel, Arup K. Ghosh, Ulf P. Trociewitz, Eric E. Hellstrom and David C. Larbalestier

*Abstract*—Bi-2212 round wire is made by the powder-in-tube technique. An unavoidable property of powder-in-tube conductors is that there is about 30% void space in the as-drawn wire. We have recently shown that the gas present in the as-drawn Bi-2212 wire agglomerates into large bubbles and that they are presently the most deleterious current limiting mechanism. By densifying short 2212 wires before reaction through cold isostatic pressing (CIPping), the void space was almost removed and the gas bubble density was reduced significantly, resulting in a doubled engineering critical current density ($J_E$) of 810 A/mm$^2$ at 5 T, 4.2 K. Here we report on densifying Bi-2212 wire by swaging, which increased $J_E$ (4.2 K, 5 T) from 486 A/mm$^2$ for as-drawn wire to 808 A/mm$^2$ for swaged wire. This result further confirms that enhancing the filament packing density is of great importance for making major $J_E$ improvement in this round-wire magnet conductor.

## I. INTRODUCTION

Bi$_2$Sr$_2$CaCu$_2$O$_x$ (Bi-2212) round wire is a very promising conductor for very high field applications [1]-[7]. It is well known that long lengths have several times lower $J_c$ than short samples and this has meant that so far no Bi-2212 user magnets have been made. The earlier studies of Shen *et al.* [8], Malagoli *et al.* [9], Kametani *et al.* [10], and Scheuerlein *et al.* [11] provide firm evidence that the large gas bubbles formed during the excursion into the melt state provide the major present current-limiting mechanism in round-wire Bi-2212 conductors. A recent study by Malagoli *et al.* [12] on the length-dependent expansion of a Bi-2212 wire after full heat treatment showed that gas pressure provokes a wire diameter expansion which increases with distance from the ends, longer samples (up to 2.4 m) often showing evident damage and leaks caused by the internal pressure. The decay of critical current density ($J_c$) away from the ends observed by Kuroda *et al.* [13] and Malagoli *et al.* [9, 12] is associated with this increase in wire diameter. The essential problem is that the final Bi-2212 powder density is only about 70% in as-drawn wire [14]. The 30% void space is normally filled with air, which is distributed uniformly on a scale much smaller than the filament diameter, but when Bi-2212 melts this well distributed porosity agglomerates into filament-sized, gas-filled bubbles, which may be filled with residual N$_2$ from the pore space, as well as additional CO$_2$ and H$_2$O from gasification of condensed C or H impurities.

Low powder packing density is a common problem for all powder-in-tube conductors such as Bi-2223 and MgB$_2$. Densification methods such as flat rolling, two axis rolling [15], [16], eccentric rolling [17], grooved rolling [18], periodic pressing [19] and cold high pressure deformation [20] have all been used for densifying Bi-2223, Bi-2212 and MgB$_2$ conductors. All these methods densify the wires, but a defect is that they do not keep the round wire shape desired by magnet designers, a shape which is also more favorable for insulation. In contrast, cold isostatic pressing (CIPping), swaging, and hot isostatic pressing (HIPping) are densification techniques that keep the round conductor shape. Overpressure processing (OP) is a modified HIPping with gas flowing for better oxygen partial pressure control [21, 22]. It has been very successful for processing Bi-2223 tapes [23-25], and is used by Sumitomo Electric for manufacturing Bi-2223 conductor [25].

Jiang *et al.* [26] recently demonstrated that $J_c$ was more than doubled by replacing residual air in the Bi-2212 filaments with pure oxygen and CIPping the wire with a pressure of 2 GPa before melt-processing in order to greatly decrease the filament void fraction. The fewer, smaller and perhaps pure O$_2$ bubbles formed in the melt allowed the critical current $I_c$ (4.2 K, 5 T) to be doubled. CIPping needs a pressure at least 65 MPa to achieve higher packing density, since the yield strength of pure silver is about 65 MPa [27], [28]. Silver alloy sheath is normally used for Bi-2212 conductors for better strength, so that pressure much higher than 65 MPa is needed for CIPping.

As shown by Miao *et al.* [29], a pressure larger than 800 MPa is preferred to achieve the high $J_c$ by CIPping. Most of the current commercial CIPping systems have a maximum pressure of 380 MPa which is not high enough for densifing Bi-2212 wire. CIPping systems with the maximum pressure higher than 800 MPa normally have a very small pressure chamber which is not feasible for densifying hundred meter

J. Jiang is with National High Magnetic Field Laboratory, Florida State University, Tallahassee, FL 32310, USA (phone: 850-645-7492; fax: 850-645-7754; e-mail: jjiang@asc.magnet.fsu.edu).
H. Miao, Y. Huang, S. Hong, and J. A. Parrell are with Oxford Superconducting Technology, Carteret, NJ 07008, USA. S. Hong is also with HJC Enterprise, 5 Badgley Drive, New Providence, NJ 07974, USA.
C. Scheuerlein is with European Organization for Nuclear Research, CH-1211 Geneva, Switzerland (email: Christian.Scheuerlein@cern.ch).
M. Di Michiel is with European Synchrotron Radiation Facility, 6 rue Jules Horowitz, F-38043 Grenoble, France.
A. K. Ghosh is with Brookhaven National Laboratory, Upton, NY 11973, USA (email: aghosh@bnl.gov).
U. P. Trociewitz, E. E. Hellstrom, and D. C. Larbalestier are with National High Magnetic Field Laboratory, Florida State University, Tallahassee, FL 32310, USA.

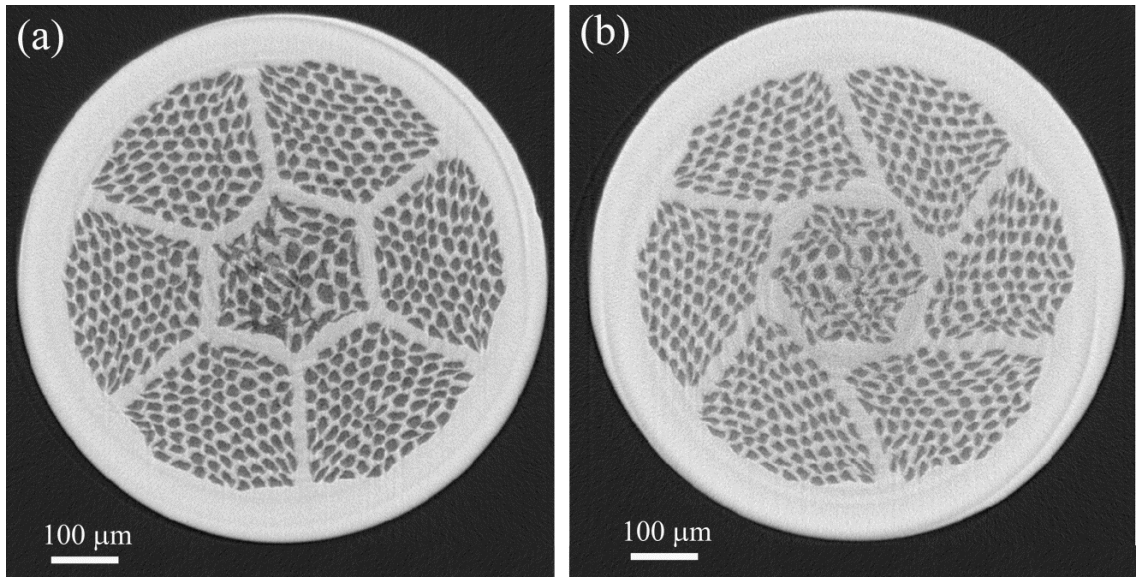

Fig.1. X-ray tomography images of transverse cross sections of (a) as-drawn wire and (b) as-swaged wire before reaction.

long Bi-2212 wire or large size magnets wound with Bi-2212 wire. Compared to CIPping, swaging is an alternative method for densifying the round wire, and it works for long length, but it is not clear how effective it is densifying Bi-2212 wire by swaging. Here we report on the effects of swaging on the gas bubble formation and critical current density of Bi-2212 round wires.

## II. EXPERIMENTAL DETAIL

The Bi-2212 wire used in this experiment was fabricated by Oxford Superconducting Technology (OST) using Nexans standard powder with a composition of $Bi_{2.17}Sr_{1.94}Ca_{0.90}Cu_{1.98}O_x$. The Bi-2212 powder was always surrounded by pure silver but the outer sheath of the 85x7 stack was made with a Ag-Mg(0.2 wt.%) alloy sheath. One 3 m long piece of the wire with 1.0 mm diameter was swaged to 0.81 mm diameter. For comparison, another 3 m long piece of the wire was drawn to 0.80 mm diameter. The diameter of 10 mm long wire sections was measured with a step of 0.1 mm along the wire axis using a Nikon VMA-2520 microscopy with a resolution of 0.1 μm. The diameter was 0.7980 ± 0.0022 mm, and 0.8099 ± 0.0034 mm, respectively, for as-drawn and swaged wires. The relative wire diameter variation was 0.28 %, and 0.42 %, respectively, for as-drawn and swaged wires. Thus, the surface smoothness of the swaged wire is comparable to that of as-drawn wire.

Straight 8 cm long pieces and 1.6 m long pieces wound on an ITER barrel were heat treated with open ends in 1 bar oxygen with a standard heat treatment (HT) schedule [26]. Samples were also quenched into brine after being heated to 887 °C (the maximum temperature of the HT) for 12 min to freeze-in the high-temperature microstructure and especially to make the bubbles formed on melting visible.

Transverse cross-sections of as-drawn, as-swaged, quenched, and fully-processed wires were dry polished using a series of SiC papers with decreasing grit sizes with final polishing conducted in a suspension of 50 nm alumina mixed with ethanol using an automatic vibratory polisher (Buehler Vibromet). Microstructures were examined with a Zeiss 1540EsB scanning electron microscope (SEM). The cross section area was measured with an Olympus BX41M-LED microscope. The wire filling factor is defined as filament cross section area divided by total wire cross section area. The filament mass density was calculated by assuming the density of silver and silver alloy to be 10.49 g/cm³. X-ray tomographic measurements were performed at the ID15A beam line of the European Synchrotron Radiation Facility (ESRF) using a monochromatic 70 keV x-ray beam with a bandwidth of 0.7 keV. The image pixel size is $1.194 \times 1.194$ μm$^2$. More details about the tomography set-up can be found in [11].

TABLE I
FILLING FACTOR AND DENSITY OF UN-REACTED WIRES, $I_c$ (4.2 K, 5 T), N VALUE (4.2 K, 5 T), $J_E$ (4.2 K,5 T), AND $J_C$ (4.2 K, 5 T) OF FULLY-PROCESSED WIRES

| Wire | Densification | Diameter | 2212 Filling factor | Mass density[a] g/cm³ | $I_c$ (4.2 K, 5 T) A | n (4.2 K, 5 T) | $J_E$ (4.2 K, 5 T) A/mm² | $J_C$ (4.2 K, 5 T) A/mm² | Reference |
|---|---|---|---|---|---|---|---|---|---|
| 85x7 | As-drawn | 0.80 mm | 0.32 | 3.66 | 244.5 | 15 | 486 | 1520 | This work |
| 85x7 | Swaged | 0.81 mm | 0.28 | 5.33 | 416.4 | 18 | 808 | 2886 | This work |
| 37x18 | CIPped | 0.78 mm | 0.22 | 5.34 | 386.4 | 20 | 809 | 3611 | [26] |

[a] Mass density of Bi-2212 filaments.

Critical currents of fully-processed wires were measured using the four-probe transport method with a 1 μV/cm criterion at 4.2 K in a magnetic field of 5 T applied perpendicular to the wire axis. The spiral sample was cut into two parts (turns 1-8, and turns 9-16) and then remounted on an $I_c$ measurement test barrel. Each 10 cm turn was measured. The overall wire critical current density $J_E$ was calculated using the whole wire cross section.

### III. RESULTS

Fig. 1 shows the x-ray tomographic images of transverse cross sections of un-reacted, as-drawn and as-swaged wires. The outer filament bundles of the swaged wire were partially twisted due to the manual feed into a regular two-die swaging machine. The main difference between these two images is that the filaments in the as-drawn wire are darker than those of the as-swaged wire. The darker shade of grey signifies a lower filament density of the as-drawn wire. The smaller filament size of the as-swaged wire is due to the densification by swaging, a result also confirmed by optical image analysis of wire cross sections and density calculations. As listed in Table 1, swaging reduced the filament region filling factor from 0.32 to 0.28, and increased the calculated Bi-2212 filament density of the wires prior to reaction from 3.66 to 5.33 g/cm$^3$. The density of the swaged wire is similar to that of CIPped wire.

Fig. 2 shows secondary electron images of transverse cross sections of the wires quenched from the melt stage (887°C/12

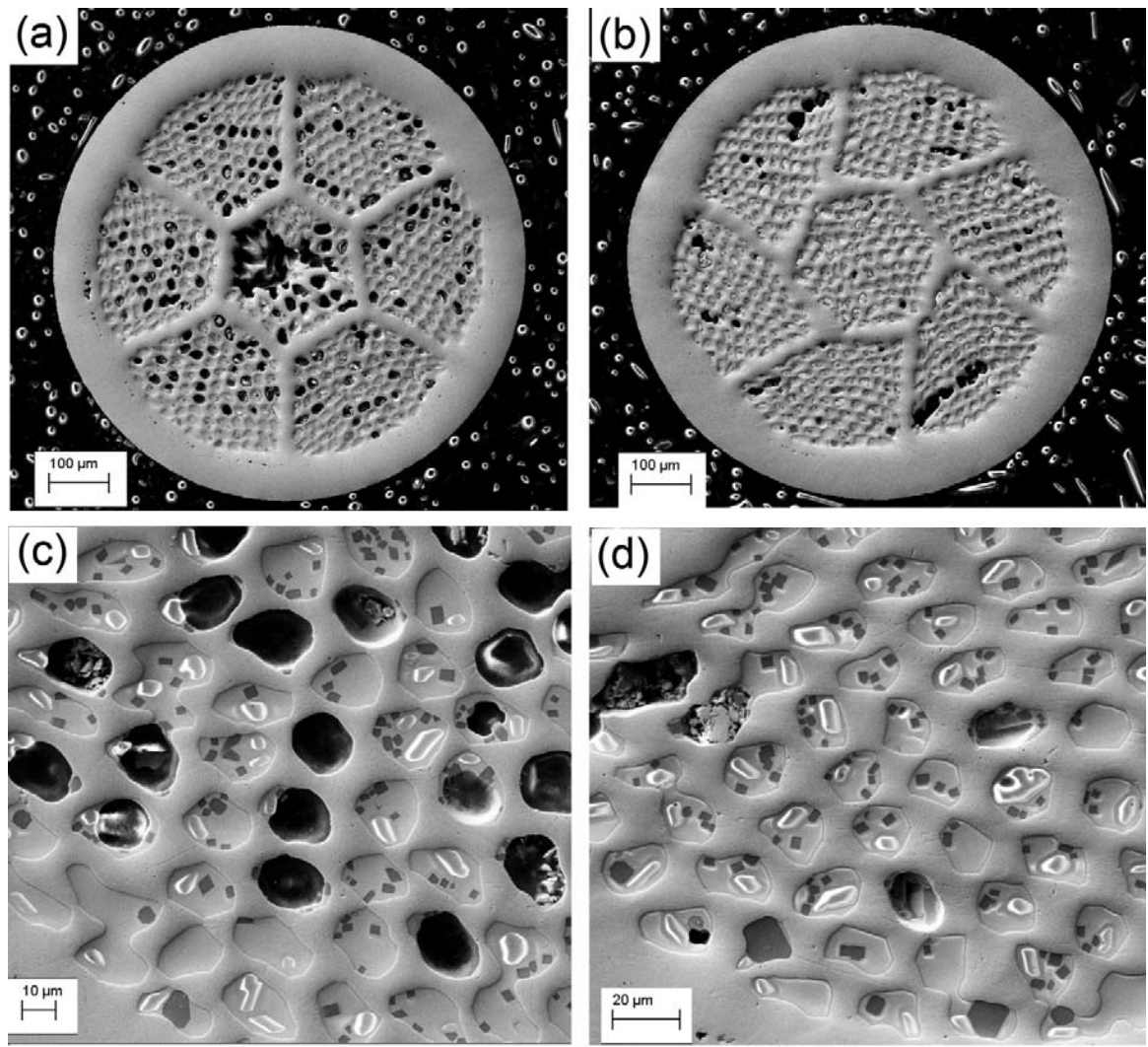

Fig.2. Scanning electron images of transverse cross sections of wires quenched from 887 C: (a) and (c) drawn wire; (b) and (d) swaged wire. The round black spots are the bubbles, the angular dark grey particles are alkaline earth cuprate $(Sr,Ca)_{14}Cu_{24}O_x$ (14:24 AEC), while the light grey regions in the filaments are liquid, and particles with white edges are the Cu-free phase $(Bi_9(Sr,Ca)_{16}O_x$, (9:16 CF). The liquid and Ag matrix are almost the same shade of grey. The gas bubbles in the quenched, as-drawn wire are generally as large as the filaments. The quenched, swaged wire had fewer gas bubbles which were generally smaller than those found in the drawn wire.

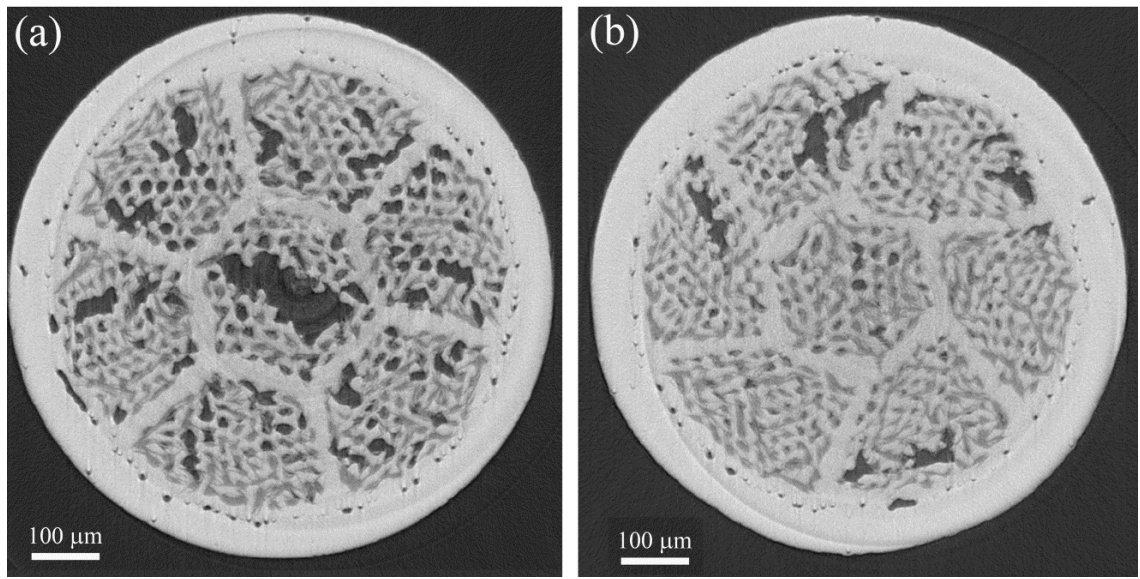

Fig.3. X-ray tomographic images of transverse cross sections of fully-processed 8 cm long wires, (a) as-drawn with $I_c$ (4.2 K, 5 T) = 244.5 A, and (b) swaged with $I_c$ (4.2 K, 5 T) = 416.4 A.

min). There were many filament-size gas bubbles in the drawn wire, and the central bundle had some very large merged bubbles, indicating very low filament density in the central bundle. In contrast, swaged wire had much fewer gas bubbles, and its central bundle was nearly bubble free. This quench study further confirmed the increased mass density and reduced bubble density in the swaged wire.

Fig. 3 shows the x-ray tomography images of fully-processed wires. The fully-processed samples showed more merged gas bubbles than the quenched samples (Fig. 2) and they are several times larger than the original filament size. This means that the gas bubbles were merging together during the melt stage of the HT throughout the time Bi-2212 solidifies. Even though swaging increased $I_c$ (4.2 K, 5 T) from 244.5 A to 416.4 A, the fully-processed, swaged wire still had some remnant gas bubbles. Thus, further $I_c$ improvement is possible. The SEM images of the fully-processed wires shown in Fig. 4 are quite typical. It is very clear that the swaged wire had many less porous filaments.

Fig. 5 shows $I_c$ (4.2 K, 5 T) as a function of position for the 1.6 m long spiral sample. The average $I_c$ (4.2 K, 5 T) is 337.5 A, which is a 38 % improvement from that of the 8 cm long as-drawn sample. The $I_c$ for turns 10-15 is more uniform than that for turns 2-8. This could be an end effect of the swaged wire. The 1.6 m long sample was cut from the 3 meter long swaged wire. Turns 2-8 could be the end of the original 3 meter long wire, and turns 10-15 the middle of the wire.

IV. DISCUSSION

Previous studies by Shen *et al.* [8], Kametani *et al.* [10] and Scheuerlein *et al.* [11] showed clearly that the gas in the wire agglomerates into filament size bubbles when Bi-2212 melts. When Bi-2212 reforms on cooling, the bubbles can be partially filled by new plate-like Bi-2212 grains, but the original bubble structure remains, even in fully-processed wires, leaving unsupported bridges of Bi-2212 shown earlier by Kametani *et al.* [10]. As shown in Figs. 2 and 3, the gas bubbles in the as-drawn wire can grow very large, with one big bubble occupying nearly one third of the central bundle in both cross sectional views of the as-drawn wire. These extremely large gas bubbles that form by merging across the silver webbing during the HT are definitely a severe current limiting mechanism. We believe that the larger the bubble, the smaller the critical current across the bubble because the current is confined to Bi-2212 grains along the Ag/bubble interface, to Bi-2212 grains that grow across such bubbles, and to filament-to-filament interconnections that may enable current flow around the bubbles [8].

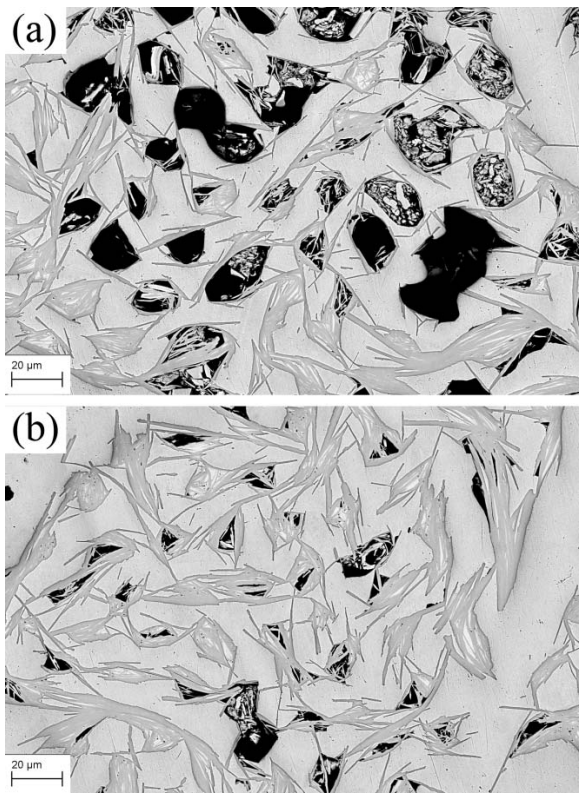

Fig.4. Scanning electron images of cross sections of fully-processed 8 cm long wires, (a) drawn wire with $I_c$(4.2 K, 5 T) = 244.5 A, and (b) swaged wire with $I_c$(4.2 K, 5 T) = 416.4 A.

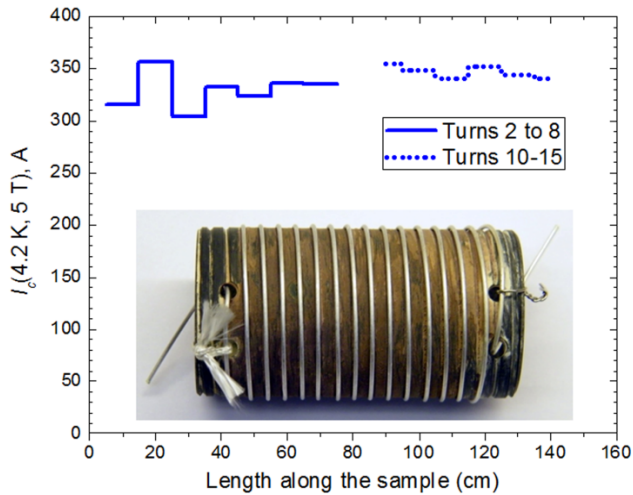

Fig.5. $I_c$ (4.2 K, 5 T) of 1.6 m long spiral of swaged wire with an average $I_c$ (4.2 K, 5 T) of 337.5 ± 32.4 A. The inset is the image of the spiral on an ITER barrel after full heat treatment.

Images of Bi-2212 are normally taken on a polished cross section, where damage from grinding and polishing (either pullout or fill in) can result in what looks like extra (or reduced) bubble area in the SEM images. By contrast, x-ray tomography is performed on the whole wire, so we have no doubt that the giant gas bubbles shown in Fig. 3 are real. The big bubble in the fully-processed, as-drawn wire also indicates that the mass density of the central bundle is quite low and carries little current compared to the outer bundles. Comparing quenched samples in Fig. 2 with fully-processed samples in Fig. 3, we can see that there are more large bubbles that formed by filament merging across the silver webbing in the fully-processed wire, indicating that extensive gas bubble merging across the web occurred when the Bi-2212 was in the melt state before the Bi-2212 reformed on cooling. This observation may also explain the strong negative correlation between critical current density and time-in-the-melt recently observed by Shen et al. [30]. Increasing the time-in-the-melt may result in more gas-bubble-driven filament merging with its consequent reduction of connectivity.

Quantitative analysis of the images in Figs. 2a and 2b shows that the area of the gas bubbles changed from 0.045 mm$^2$ in quenched, as-drawn wire to 0.011 mm$^2$ in quenched, swaged wire, i.e. a 75% bubble area reduction by swaging. The filament density values listed in Table 1 are 3.66 and 5.33 g/cm$^3$ for un-reacted as-drawn and as-swaged wires, respectively, corresponding to an increase in the relative Bi-2212 density from 55% to 82%. Swaging achieved a similar filamnet density to CIPping (Table 1) and they attain almost identical $J_E$ (4.2 K, 5 T). However, Figs. 3 and 4 show that there are still some large gas bubbles in the fully-processed, swaged wire, so further $J_c$ improvement is highly possible.

Previous work by Malagoli et al. [9] on a 1 m long 37x18 wire with open ends showed that $I_c$ dropped 25% from end to middle, which was correlated to the wire diameter expansion caused by internal gas pressure [12]. Compared to the result of previous work [9], Fig. 5 shows that swaging not only increased the wire $I_c$, but also increased $I_c$ uniformity along the length. The fact that the ends of the fully-reacted 1.6 m long spiral are clean and that the $I_c$ is relatively uniform along the length, indicate lower internal gas pressure during the HT in the 1.6 m long swaged wire compared to the 1 m long as-drawn wire [9, 12]. One reason might be that swaging can "push" residual gas, presumably air, out of the back end of the wire. The internal gas pressure produced during HT comes from the gas in the pore space and also from the gasification of condensed phases such as carbon and hydrogen impurities in the filaments. Since swaging reduced the pore space and also depressed the internal gas pressure effect in the 1.6 m long spiral, it seems that the residual gas in the pore space is the main source of the internal gas pressure. Otherwise, if the carbon and hydrogen impurities are the main gas source, we would expect that the extra gas in the reduced pore space, from the gasification, could expand the swaged wire seriously during the HT, resulting in lower critical current.

Supposing that the gas inside the wire is air (80% $N_2$ and 20% $O_2$), and that the total gas pressure inside the wire at room temperature (about 300 K) is 1 bar, according to the ideal gas law the total pressure inside the wire will become about 4 bar when the wire is heated to the maximum HT temperature 890 $^o$C (1163 K). Since oxygen can diffuse through the silver sheath, the main internal pressure comes from the $N_2$, which is about 3.2 bar at 890 $^o$C. Our recent overpressure experiments indeed show that a pressure of 5 bar can limit the wire expansion during the HT [31], indicating that the internal pressure may be about 5 bar, which is comparable to the 3 bar partial pressure of $N_2$. Thus, we believe that the internal gas pressure contributed from the gasification of the condensed phases such as carbon and hydrogen impurities is limited.

Even though swaging did not achieve full densification, it improved the wire $I_c$ significantly. In principle, one could make long-length, high-packing-density wire with a better precision swaging machine.

## V. CONCLUSION

We found a significant improvement of $J_c$ in a Bi-2212 round wire with 85x7 filaments by swaging before melt-processing. Like CIPping [26], swaging densified the Bi-2212 filaments, resulting in much fewer gas bubbles in the melt and nearly doubling $I_c$ (4.2 K, 5 T). $J_E$ (4.2 K, 5 T) increased to 808 A/mm$^2$, and $J_c$ (4.2 K, 5 T) to 2886 A/mm$^2$. This result further confirms that gas bubbles are the major current limiting mechanism in present Bi-2212 round wires and that enhancing the filament packing density is of great importance for making major $J_c$ improvements.

ACKNOWLEDGMENT

We are very grateful for discussions with F. Kametani, M. Dalban-Canassy, and P. J. Lee at FSU, with T. Shen at FNAL, and with members of the Very High Field Superconducting Magnet Collaboration. Technical support from M. Hannion, W. L. Starch, V. S. Griffin and J. Craft is also gratefully acknowledged. The work at the NHMFL was supported by an ARRA grant from the US Department of Energy Office of High Energy Physics and by the National High Magnetic Field Laboratory, which is supported by the National Science Foundation under NSF/DMR-0654118, and by the State of Florida. We acknowledge the ESRF for beam time on ID15A.